\documentclass[prl,nofootinbib,preprint,superscriptaddress]{revtex4}

\usepackage{graphicx}
\usepackage{amsmath}
\usepackage{amsfonts}
\usepackage{amssymb}
\usepackage{dsfont}
\usepackage{dcolumn}
\usepackage{bm}
\usepackage{axodraw4j}
\usepackage{pstricks}
\usepackage{color}

\newcommand{\beqn}{\begin{eqnarray}}
\newcommand{\eeqn}{\end{eqnarray}}
\newcommand{\be}{\begin{equation}}
\newcommand{\ee}{\end{equation}}

\newcommand{\mathsym}[1]{{}}

\def \cha{\tilde{\chi}^{\pm}_1}

\def \na{\tilde{\chi}^{0}_1}
\def \nb{\tilde{\chi}^{0}_2}

\def \ta{\tilde{t}_1}

\def \sta{\tilde{\tau}_1}

\begin{document}

\title{
SUSY and Higgs Signatures Implied by Cancellations in  $b\to s\gamma$}
\author{Ning Chen}
\affiliation{C.N.\ Yang Institute for Theoretical Physics, 
Stony Brook University, Stony Brook, NY 11794, USA}
  
\author{Daniel Feldman}
\affiliation{Michigan Center for Theoretical Physics,
University of Michigan, Ann Arbor, MI 48109, USA}

\author{Zuowei Liu}
\affiliation{C.N.\ Yang Institute for Theoretical Physics, 
Stony Brook University, Stony Brook, NY 11794, USA}

\author{Pran Nath}
\affiliation{Department of Physics, Northeastern University,
 Boston, MA 02115, USA}


\preprint{YITP-SB-09-33; MCTP-09-52; NUB-TH-3264}


\begin{abstract}
Recent re-evaluations of  the Standard Model (SM) contribution to 
${\mathcal Br}(b\to s\gamma)$
 hint at a positive correction from new physics. 
 Since a charged Higgs boson exchange always gives a positive contribution to this branching ratio, the constraint
 points to the possibility of a relatively light charged Higgs.
It is found that under the HFAG constraints and with re-evaluated SM results 
large cancellations between the charged Higgs and the chargino contributions in 
supersymmetric models occur. 
Such cancellations then correlate the charged Higgs and the chargino masses often implying
both are light. 
Inclusion of the more recent evaluation of $g_{\mu}-2$ is also considered. The combined constraints
 imply the existence of several light sparticles. Signatures  arising from these light sparticles  are
investigated and the analysis indicates the possibility of their early discovery at the LHC
in a significant part of the parameter space. 
We also show that for certain restricted regions of the parameter space, such as for very 
large $\tan\beta$ under the $1\sigma$ HFAG constraints,
the signatures from Higgs production supersede those from sparticle production and
 may become the primary signatures for the discovery of supersymmetry.

 \end{abstract}

\maketitle


{\bf Introduction}:
Recently a re-evaluation of the SM result for the  branching ratio for the 
flavor changing neutral current (FCNC) process  $b\to s\gamma$ 
including NNLO corrections in QCD has been given   \cite{Misiak:2006zs} 
${\mathcal Br}(b\rightarrow s\gamma) =(3.15\pm 0.23) \times 10^{-4}$. 
This new estimate lies lower than the current 
experimental value which is given by the Heavy Flavor Averaging Group 
(HFAG)   \cite{Barberio:2008fa} along with the BABAR, 
Belle and CLEO  experimental results: 
${\mathcal Br}(B \to X_s \gamma) =(352\pm 23\pm 9) \times 10^{-6}$.
The above result hints at a positive contribution to this process arising from new physics.
It is known from the early days that the experimental value of  the branching ratio $b\to s\gamma$ 
is a very strong
constraint on  the parameter space of most classes of SUSY models  \cite{bbmr,Nath:1994tn} 
(for more recent theoretical evaluations of 
${\mathcal Br}(b\to s\gamma)$ in supersymmetry see   \cite{susybsgamma}).  
A positive contribution to ${\mathcal Br}(b\to s\gamma)$ implies either the existence of a light
charged Higgs  exchange which always gives a positive contribution  \cite{Hewett:1992is}  or the 
existence of a  light chargino which can give  either a positive or
a  negative contribution  \cite{Garisto:1993jc}. 
 A  significant cancellation between the  charged  Higgs loop contribution and  the
  chargino contribution
  implies that individual contributions from  
the charged  Higgs  loops and the gaugino loops must each be often multiples of 
their sum. Such cancellations  then necessarily imply that some of the sparticles
that enter in the supersymmetric contributions to the FCNC loops must be  relatively
light and thus should be  accessible in early runs  at the LHC.

In addition to the above, recently  the  difference between experiment and 
the standard model prediction of the anomalous magnetic moment of the muon,  
$a_{\mu}=(g_{\mu}-2)/2$ seem to converge  \cite{Davier:2009zi}
towards roughly a $3\sigma$  deviation from the SM value. Thus the most recent analysis gives 
$\delta a_{\mu}= a_{\mu}^{exp}-a_{\mu}^{SM}$ as   \cite{Davier:2009zi} 
$\delta a_{\mu} = (24.6\pm 8.0)\times 10^{-10}$. 
It is well known that  supersymmetric electroweak contributions to $g_{\mu}-2$ can be 
as large or larger than the SM electroweak corrections  \cite{yuan}. Further,  a large   deviation
of $g_{\mu}-2$ from the SM is a harbinger  \cite{Czarnecki:2001pv},
for the observation of low lying sparticles  \cite{Chattopadhyay:2001vx,Everett:2001tq,fm}  
at colliders with the experimental data  putting   upper limits on some of the sparticle 
masses in SUGRA models  \cite{Chattopadhyay:2001vx}.
The positive correction to $b\to s\gamma$ which is of size $(1-1.5)\sigma$ together with  
the $3\sigma$ level deviation of $g_{\mu}-2$ from the standard model value points to the existence 
of some of the sparticles being light.

\begin{figure}[t]
\includegraphics[width=7cm,height=6cm]{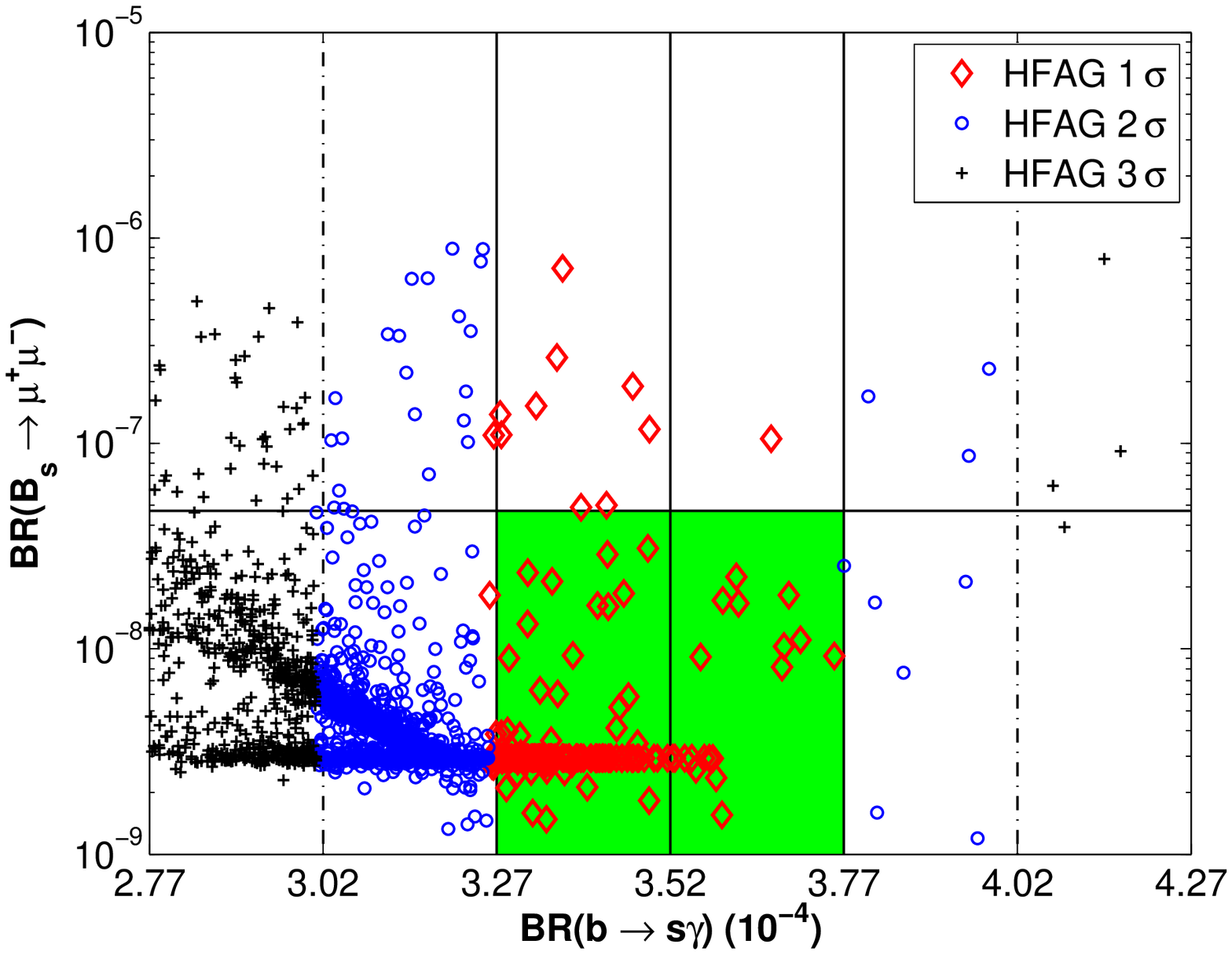}
\includegraphics[width=7cm,height=6cm]{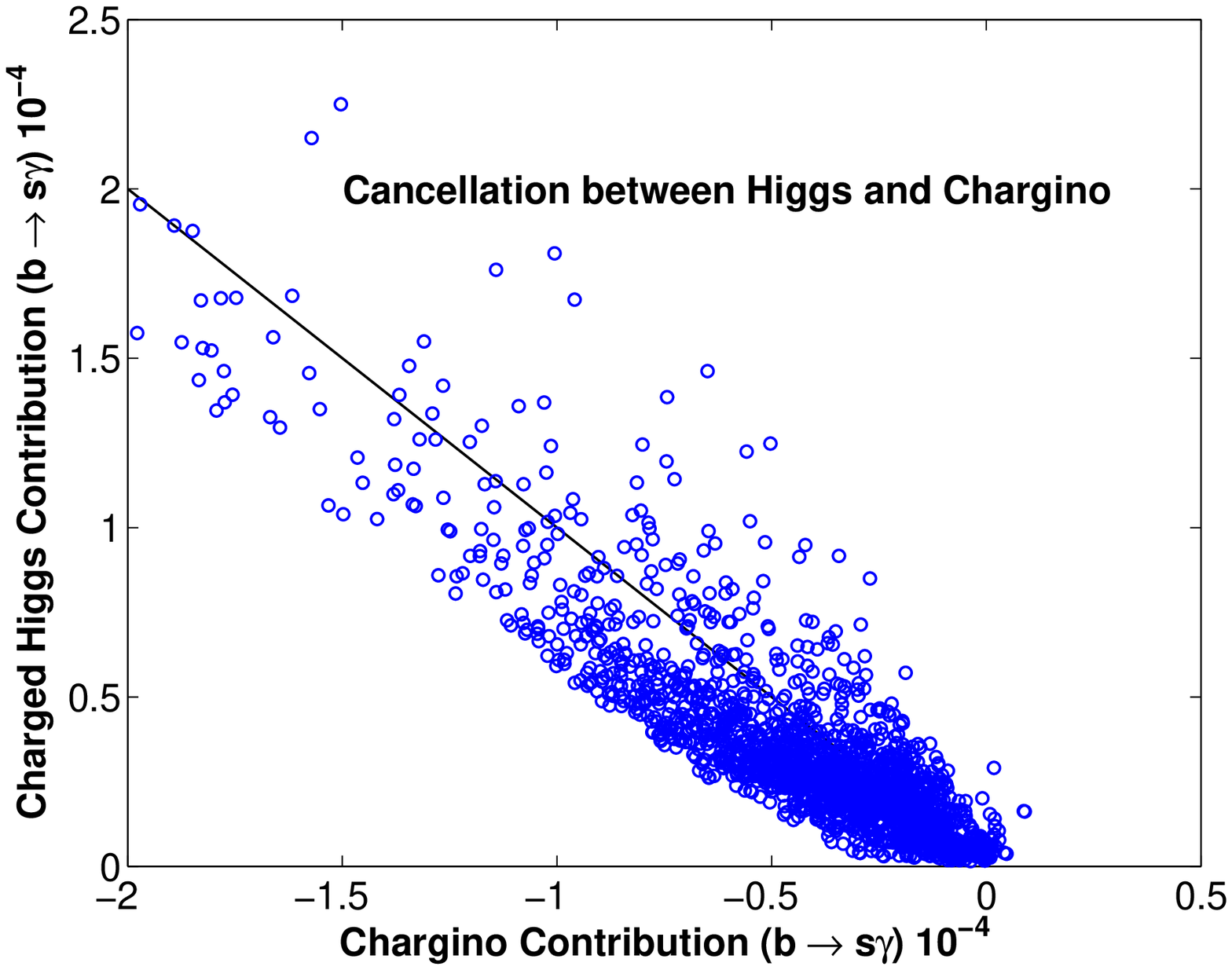}
\caption{
Left: The shaded region are the models 
that survive the constraint ${\mathcal Br}(B_s\to \mu^+\mu^-)<4.7\times 10^{-8}$
and the   constraint ${\mathcal Br}(b\to s\gamma)=3.52\pm 0.25$ as given by HFAG.
Right: Charged Higgs contribution vs the chargino contribution. 
One finds that in most models 
the chargino exchange contributions are almost always negative and are strongly
correlated  with the charged Higgs contributions.
The individual contributions from the charged Higgs and the  
chargino are computed using SusyBSG  \cite{Degrassi:2007kj}.
} 
\label{d}
\end{figure}

\noindent
{\bf Analysis}:
 In this work we investigate the implications of  
 the revised constraints in the framework of supergravity grand unified models  \cite{msugra}  
following the analysis of   \cite{Feldman:2007zn} 
with the parameter space 
characterized by 
parameters $m_0, m_{1/2}, A_0, \tan\beta$, sign($\mu$) 
where for  Monte Carlo simulations we have assumed the following range: 
$m_0<4$ TeV, $m_{1/2}<2$ TeV, $|A_0/m_0|<10$, 
and $1<\tan\beta<60$ with $\mu>0$ for three million candidate models. 
For the purpose of selecting viable models from the large scan, 
we impose the following set of constraints: 
(conservative bounds are given here 
to illustrate the constraining effects and also to account 
for experimental and theoretical uncertainties)
(i) The 5-year WMAP data constrains the relic density of dark matter 
so that  
$\Omega_{\rm DM} h^2 = 0.1131\pm 0.0034$  \cite{Komatsu:2008hk}. 
The  bound $ 0.0855<\Omega_{\tilde \chi_1^0} h^2<0.1189$  
  \cite{Spergel:2006hy} is taken; 
(ii) A $3$ $\sigma$ constraint for $b\to s\gamma$ is taken around the HFAG value
(a stricter constraint will be considered later); 
(iii) The 95\% (90\%) C.L. limit reported  by CDF in 
${\mathcal Br}( B_s \to \mu^{+}\mu^{-})$ is  
$5.8 \times 10^{-8}$ ($4.7 \times 10^{-8}$)  \cite{2007kv} 
(we take ${\mathcal Br}( B_s \to \mu^{+}\mu^{-}) < 10^{-6}$);
(iv) $\delta a_{\mu} \in  (-5.7,47)\times 10^{-10}$ is taken as in 
  \cite{Djouadi:2006be}
(a stricter limit on $\delta a_{\mu}$ will be discussed in the last section);
(v) The following mass limits on light Higgs boson mass and on sparticle masses are
imposed:
$m_h>100 ~{\rm GeV}$, (the current data sets limits for the MSSM case of
$m_h> 93$ GeV at  95\% C.L.  \cite{Dawson:2005vi,lephiggs}) 
$m_{\cha}>104.5 ~{\rm GeV}$, 
$m_{\ta}>101.5 ~{\rm GeV}$, 
$m_{\sta}>98.8 ~{\rm  GeV}$,
where $h,\tilde  \chi_1^{\pm}, \tilde t_1, \tilde \tau_1$ are the lightest Higgs boson,
the chargino, the stop and the stau. 
For the calculations of the relic density of $\tilde \chi_1^0$, 
we use MicrOMEGAs  \cite{MICRO} with sparticle and Higgs masses 
calculated by the RGE package SuSpect  \cite{SUSPECT}.  
Evaluation of the branching ratio $b\to s\gamma$ has been carried out 
with both MicrOMEGAs and SusyBSG  \cite{Degrassi:2007kj}.
The models that pass the above constraints are exhibited in Fig.(\ref{d}).

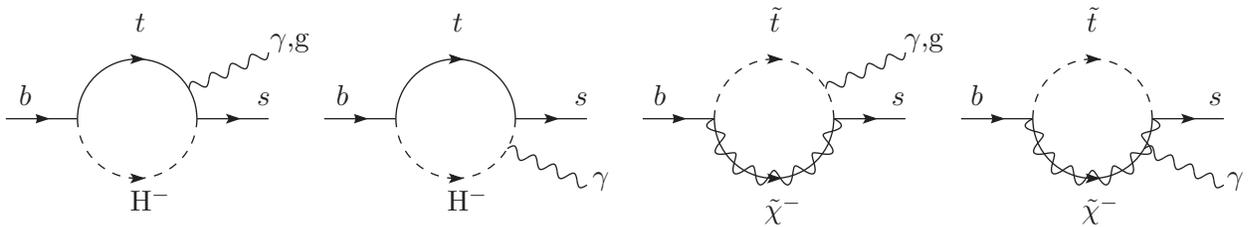
\begin{figure}
\begin{center}
\fcolorbox{white}{white}{
  \begin{picture}(480,100) (0,-50)
    \SetWidth{0.5}
    \SetColor{Black}
    \Line[arrow,arrowpos=0.5,arrowlength=3,arrowwidth=1,arrowinset=0.1](0,0)(27,0)
    \Arc[arrow,arrowpos=0.5,arrowlength=3,arrowwidth=1,arrowinset=0.2, clock](49.5,0)(22.5,-180,-360)
    \Arc[dash,dashsize=3,arrow,arrowpos=0.5,arrowlength=3,arrowwidth=1,arrowinset=0.2](49.5,0)(22.5,-180,0)
    \Line[arrow,arrowpos=0.5,arrowlength=3,arrowwidth=1,arrowinset=0.2](72,0)(99,0)
    \Photon(69,11.25)(99,25){2}{4}
    \Text(5,5)[lb]{{\Black{$b$}}}
    \Text(49,33)[lb]{{\Black{$t$}}}
    \Text(47,-35)[lb]{{\Black{$\rm H^-$}}}
    \Text(95,5)[lb]{{\Black{$s$}}}
    \Text(100,25)[lb]{{\Black{$\gamma$,g}}}   
    \Line[arrow,arrowpos=0.5,arrowlength=3,arrowwidth=1,arrowinset=0.1](120,0)(147,0)
    \Arc[arrow,arrowpos=0.5,arrowlength=3,arrowwidth=1,arrowinset=0.2, clock](169.5,0)(22.5,-180,-360)
    \Arc[dash,dashsize=3,arrow,arrowpos=0.5,arrowlength=3,arrowwidth=1,arrowinset=0.2](169.5,0)(22.5,-180,0)
    \Line[arrow,arrowpos=0.5,arrowlength=3,arrowwidth=1,arrowinset=0.2](192,0)(219,0)
    \Photon(189,-11.25)(219,-25){2}{4}
    \Text(125,5)[lb]{{\Black{$b$}}}
    \Text(169,33)[lb]{{\Black{$t$}}}
    \Text(167,-35)[lb]{{\Black{$\rm H^-$}}}
    \Text(215,5)[lb]{{\Black{$s$}}}
    \Text(222,-27)[lb]{{\Black{$\gamma$}}}
    \Line[arrow,arrowpos=0.5,arrowlength=3,arrowwidth=1,arrowinset=0.1](240,0)(267,0)
    \Arc[dash,dashsize=3,arrow,arrowpos=0.5,arrowlength=3,arrowwidth=1,arrowinset=0.2, clock](289.5,0)(22.5,-180,-360)
    \Arc[arrow,arrowpos=0.5,arrowlength=3,arrowwidth=1,arrowinset=0.2](289.5,0)(22.5,-180,0)
    \PhotonArc(289.5,0)(22.5,-180,0){3}{7.5}
    \Line[arrow,arrowpos=0.5,arrowlength=3,arrowwidth=1,arrowinset=0.2](312,0)(339,0)
    \Photon(309,11.25)(339,25){2}{4}
    \Text(245,5)[lb]{{\Black{$b$}}}
    \Text(289,33)[lb]{{\Black{$\tilde t$}}}
    \Text(287,-40)[lb]{{\Black{$\rm \tilde\chi^-$}}}
    \Text(335,5)[lb]{{\Black{$s$}}}
    \Text(340,25)[lb]{{\Black{$\gamma$,g}}}
    \Line[arrow,arrowpos=0.5,arrowlength=3,arrowwidth=1,arrowinset=0.1](360,0)(387,0)
    \Arc[dash,dashsize=3,arrow,arrowpos=0.5,arrowlength=3,arrowwidth=1,arrowinset=0.2, clock](409.5,0)(22.5,-180,-360)
    \Arc[arrow,arrowpos=0.5,arrowlength=3,arrowwidth=1,arrowinset=0.2](409.5,0)(22.5,-180,0)
    \PhotonArc(409.5,0)(22.5,-180,0){3}{7.5}
    \Line[arrow,arrowpos=0.5,arrowlength=3,arrowwidth=1,arrowinset=0.2](432,0)(459,0)
    \Photon(429,-11.25)(459,-25){2}{4}
    \Text(365,5)[lb]{{\Black{$b$}}}
    \Text(409,33)[lb]{{\Black{$\tilde t$}}}
    \Text(407,-40)[lb]{{\Black{$\rm \tilde\chi^-$}}}
    \Text(455,5)[lb]{{\Black{$s$}}}
    \Text(462,-27)[lb]{{\Black{$\gamma$}}}
  \end{picture}
}
\end{center}
\caption{Leading order contributions to $b\to s\gamma$ from charged Higgs and chargino 
loops in supersymmetry. 
}
\label{fig:b2sgfeyn}
\end{figure}

{\bf Cancellation of  charged Higgs and chargino loop contributions to  ${\mathcal Br}(b\to s\gamma)$}:\\
We discuss now in further detail the cancellation between the charged Higgs and the chargino loop
contributions in the process $b\to s\gamma$ and the implications of this cancellation, 
which may point to a light charged  Higgs mass. 
The effective interaction that controls the $b\to s\gamma$ decay is given by
\beqn
H_{\rm eff}= - 2\sqrt 2 G_F V^*_{ts}V_{tb} \sum_{i=1}^8 C_i(Q) O_i(Q),
\eeqn
where $V_{ts}, V_{tb}$ are the CKM matrix elements, $O_i(Q)$ are the effective dimension six operators and 
$C_i(Q)$ are the Wilson coefficients and $Q$ is the renormalization group scale. 
The $b\to s\gamma$ receives contributions only from $C_2, C_7, C_8$ where 
the corresponding operators are $O_2=(\bar c_L\gamma^{\mu} b_L) (\bar s_L \gamma_{\mu} c_L)$, 
$O_7=(e/16\pi^2) m_b(\bar s_L\sigma^{\mu\nu} b_R)F_{\mu\nu}$, and $O_8 =(g_s/16\pi^2) 
m_b(\bar s_L \sigma^{\mu\nu} T^a b_R) G^a_{\mu\nu}$. The dominant contribution arises from 
$C_7$, where to leading order $C_7(m_b)$ is given by 
\beqn
C_7^{(0)}(m_b) = \eta^{16/23} C_{7}(M_W) +\frac{8}{3} (\eta^{16/23} -  \eta^{14/23}) C_{8}(M_W) + C
\eeqn
and where $\eta= \alpha_s(M_W)/\alpha_s(Q_b)$ and C ($\simeq .175$) arises from operator mixing. 
   Now $C_{7,8}$ contain the standard model and new physics contributions  so that 
   \beqn
   C_{7,8}(M_W)= C_{7,8}^{W} (M_W)+ C_{7,8}^{H}(M_W)+ C_{7,8}^{\chi}(M_W).
   \label{3}
   \eeqn
Here $C_{7,8}^{W}$ is the standard  model contribution arising from the W boson exchange, 
$C_{7,8}^{H}$ is the supersymmetric contribution from the charged Higgs exchange 
and $C_{7,8}^{\chi}$ is the contribution from the chargino exchange (see Fig(\ref{fig:b2sgfeyn})).
In addition to the constraints on models arising from the 
 ${\mathcal Br}(b\to s\gamma)$ experiment, there are  also constraints from
 the    ${\mathcal Br}(B_s\to \mu^+\mu^-)$ experiment. In the left panel of Fig.(\ref{d}) we display the theoretical
 predictions  in the ${\mathcal Br}(B_s\to \mu^+\mu^-) - {\mathcal Br}(b\to s \gamma)$ plane, where the
 $1\sigma$, $2\sigma$, $3\sigma$ corridors around the HFAG value of ${ \mathcal Br}(b\to s \gamma)$ 
 are also exhibited.  The analysis of  the left panel Fig.(\ref{d}) exhibits that the parameter space gets reduced
 in a significant way as the $ {\mathcal Br}(b\to s \gamma)$  constraint becomes more stringent.  
 We now note that the sign of the chargino contribution $C_{7,8}^{\chi}$  in Eq.(\ref{3}) 
has a very dramatic effect  on the size of the supersymmetric contribution.  
A positive contribution would add constructively with the charged Higgs contribution $C_{7,8}^{H}$ 
while a negative contribution cancels partially the charged Higgs contribution reducing significantly the overall
size. A numerical analysis shows that essentially for  all the model points that lie in the $3\sigma$ corridor around
the HFAG value the chargino contribution is negative and often large resulting in large cancellations. 
We exhibit this in the  right panel of Fig.(\ref{d}). 
One finds that a majority of the models are clustered around the standard model prediction 
of the $b\to s\gamma$. As discussed above this is a consequence of the {\it cancellation} 
between the charged Higgs and the chargino loop diagrams.

In the cancellations  discussed above,  the individual contributions from the
charged Higgs loop and from the  chargino loop are often much larger than the total 
SUSY contribution as exhibited in the right panel of Fig.(\ref{d}). 
This implies that some of the sparticle spectrum must be light to allow
for such large individual contributions in the branching ratio $b\to s\gamma$. 
The above also indicates that if the chargino is light, then correspondingly the charged
Higgs must be correspondingly light to generate a large compensating contribution.
So the cancellation phenomenon then strongly correlates  the charged Higgs mass and 
the chargino mass in the region of large cancellations, i.e., in the region where the magnitude
of the loop contributions from the chargino and from the charged Higgs are individually 
multiples of their sum.

\begin{figure}[h]
\includegraphics[width=7cm,height=6cm]{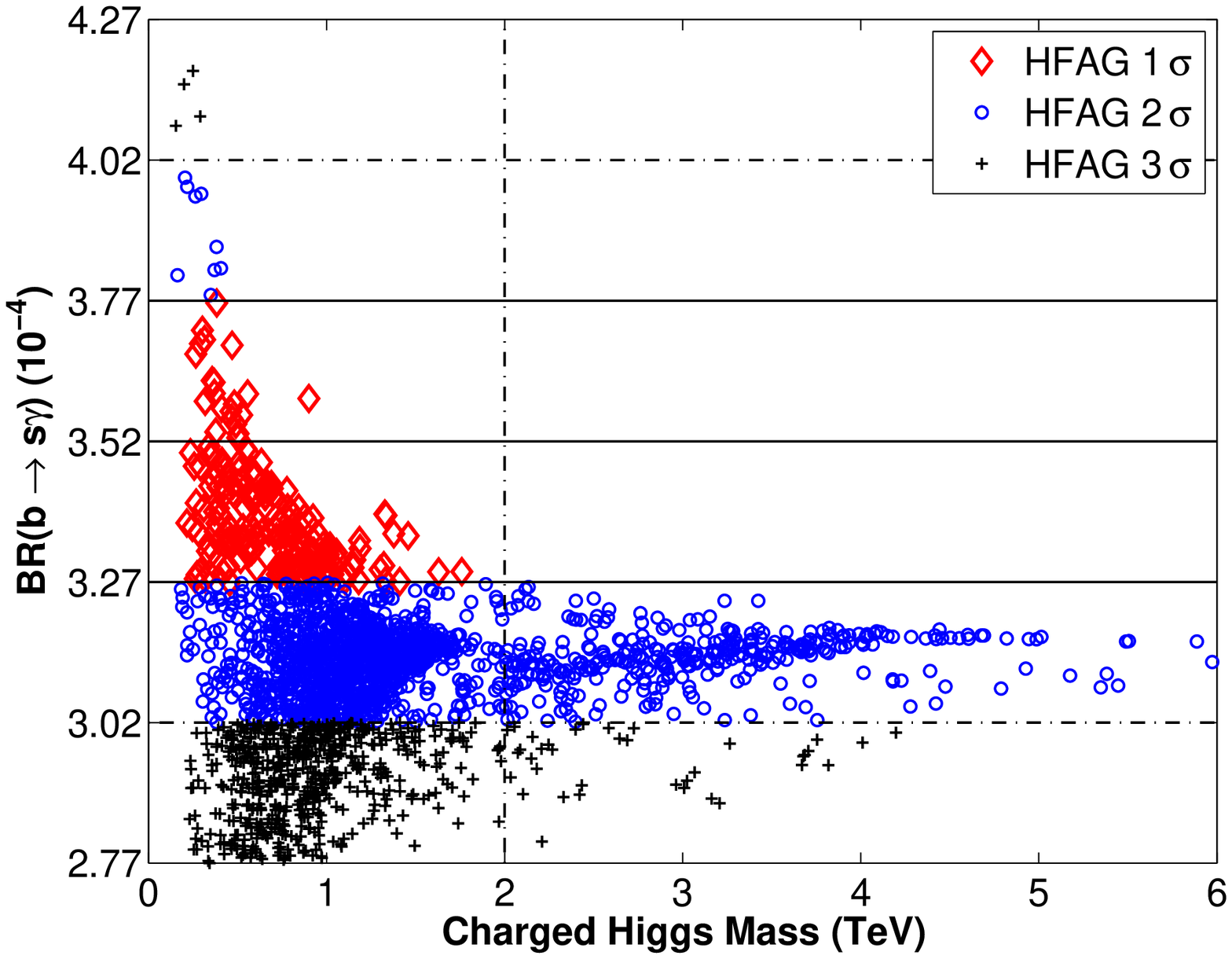}
\includegraphics[width=7cm,height=6cm]{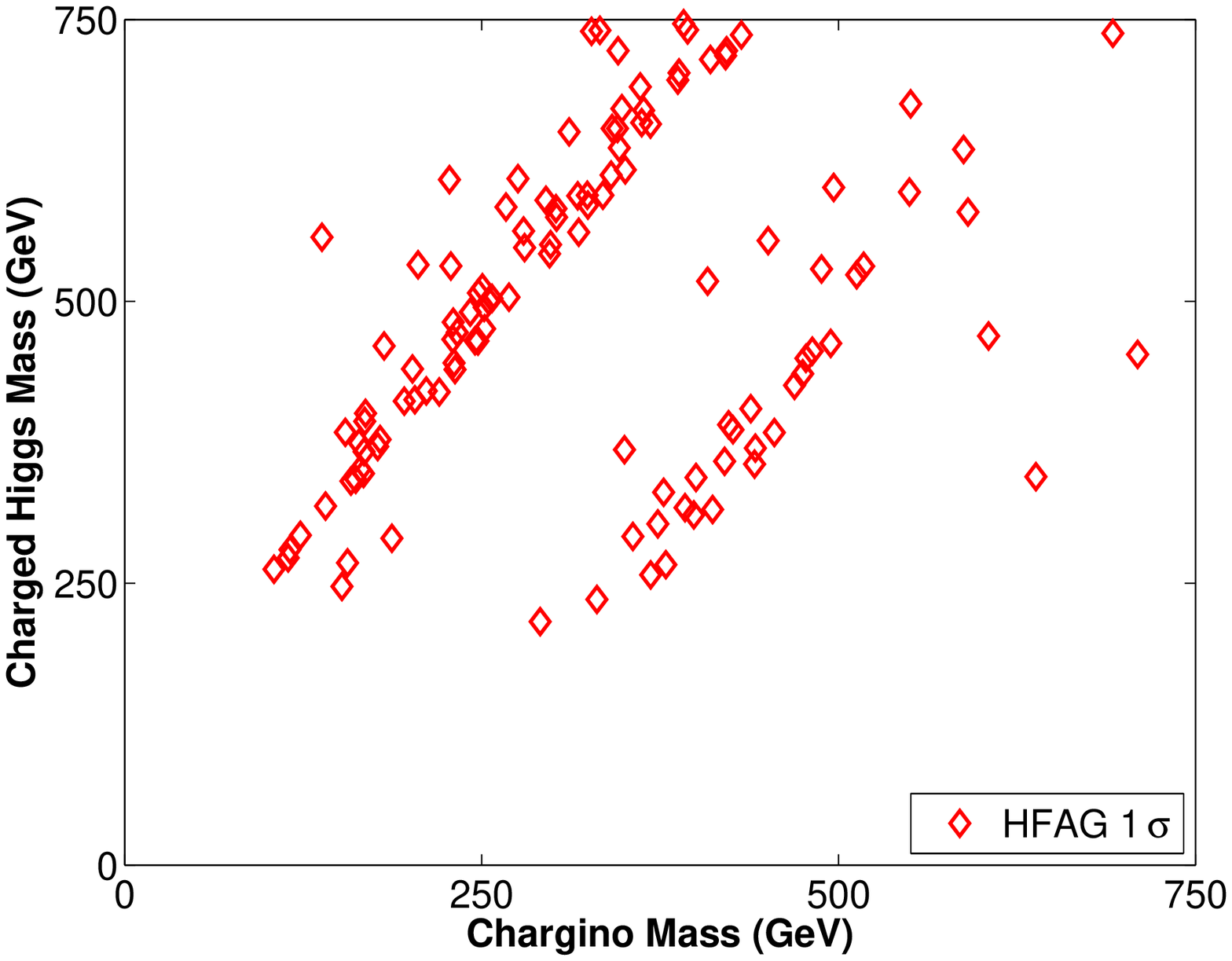}
\caption{
Left: A display of the correlation between ${\mathcal Br}(b\to s\gamma)$ 
and the charged Higgs boson mass showing the relative lightness of 
the charged Higgs boson mass in the $1\sigma$, $2\sigma$ and $3\sigma$ 
 corridors around the HFAG value. 
Right: A display of the model points  in
the charged Higgs  mass vs the light chargino mass plane 
within the $1\sigma$ corridor around 
the HFAG value  
in a large  portion of the parameter space.
} 
\label{fig:mass}
\end{figure}

\begin{figure}[h]
\includegraphics[width=7cm,height=6cm]{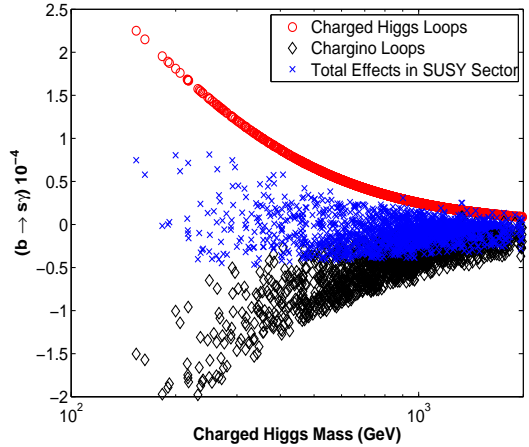}
\caption{
A display of the contributions from the charged Higgs loop, the chargino loop 
(and also other gaugino loops), and the total effect beyond the SM. }
\label{fig:cancel}
\end{figure}

An illustration of the correlation between the charged Higgs mass and ${\mathcal Br}(b\to s\gamma)$
is given in the left panel of Fig.(\ref{fig:mass}) in $1\sigma, 2\sigma, 3\sigma$ corridors around
the HFAG value. The analysis shows that a more stringent ${\mathcal Br}(b\to s\gamma)$ constraint
 typically leads to a lighter charged Higgs mass. Further, as stated earlier the cancellation phenomenon also 
 correlates  the chargino mass to the charged Higgs mass. This is illustrated the right panel of Fig.(\ref{fig:mass}).
 Specifically, here one finds that for the model points 
within HFAG  $1\sigma$, a light charged Higgs mass often requires a light 
chargino mass  to cancel the loop.
So one expects to have light Higgs and a light chargino with comparable sizes. 
The cancellation between the charged Higgs  contribution and the 
chargino contribution is also shown in Fig.(\ref{fig:cancel}) where 
the models with charged Higgs mass below $2$ TeV are plotted. 
The charged Higgs contribution increases with decreasing charged 
Higgs mass, which forces the chargino contribution to increase 
in magnitude with decreasing charged Higgs mass in order 
that the total effect is consistent with the HFAG constraints. 
We note that in the Two Higgs Doublet Model (THDM), the charged Higgs mass is also 
constrained from below, since there is no gaugino contributions to cancel the large 
positive contribution from the light charged Higgs. Thus, under the same constraints, 
the allowed charged Higgs mass can be much smaller in SUGRA models with a 
MSSM spectrum than in the THDM.  
We also note that the ${\mathcal Br}( B_s \to \mu^{+}\mu^{-})$ 
constraint becomes important for the MSSM with large $\tan\beta$. 
The current experimental limit imposes a lower bound on the Higgs mass for models with
large $\tan \beta$.

\begin{figure}[h]
\includegraphics[width=7cm,height=6cm]{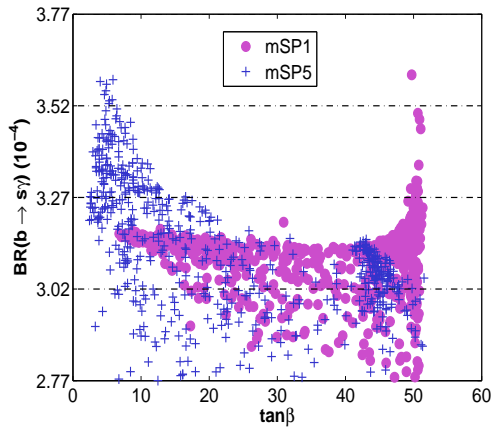}
\caption{
  $b\to s\gamma$ vs $\tan\beta$ for the mass patterns 
mSP1 and mSP5. 
The analysis of the figure shows that the 
$1\sigma$ $b\to s\gamma$ constraint selects models in distinct regions
of  $\tan\beta$: (i) a region of low $\tan\beta$  where the allowed  models
are mostly of type mSP5, and (ii) a region of large $\tan\beta$ where the
allowed the models are mostly of type mSP1. }
 \label{fig:tb}
\end{figure}

A display of the ${\mathcal Br}(b\to s\gamma)$  vs $\tan\beta$ for mSP1 and 
mSP5 models\footnote{ mSPs are supergravity  mass hierarchies
as defined in earlier 
works  \cite{Feldman:2007zn}, where
(mSP1,mSP5) have a (chargino, stau) NLSP
respectively .
}
is given in Fig.(\ref{fig:tb}) and the models that pass the 1$\sigma$ 
corridor cut on ${\mathcal Br}(b\to s\gamma)$
around the experimental value are shown. One finds that in the region of 
the 1$\sigma$ HFAG  corridor, the models from mSP1 where the lighter chargino is the NLSP have 
large $\tan\beta$ values around 50, while the models from mSP5 where the lighter stau 
is the NLSP has much smaller $\tan\beta$ values. 
We therefore collectively refer to models that reside in the $\tan\beta$ region 
where $\tan\beta < 40$ as low and high $\tan\beta$ models, ``LH $\tan\beta$ models''. 
We segregate these LH  models from those in which $\tan\beta\geq 40$ denoting these as
very high $\tan\beta$ models, ``VH $\tan\beta$ models'',  
for all the models that fall within the  $1\sigma$ corridor around the HFAG value.
We do so for all the different mass hierarchical patterns 
with mSP1 and mSP5 serving as illustrative examples. 
Typically the ``LH $\tan\beta$ models'' are the ones in which the stau, the stop, or the gluino
can be light, while the ``VH $\tan\beta$ models'' are the ones where  the chargino, 
or the Higgs is the next heavier particle than the LSP.
Some implications of the  
updated constraints on Higgs masses are  also given in   \cite{Feldman:2007fq,Barenboim:2007sk,Dudley:2009zi}.

\begin{figure}[h]
\includegraphics[width=16cm,height=7cm]{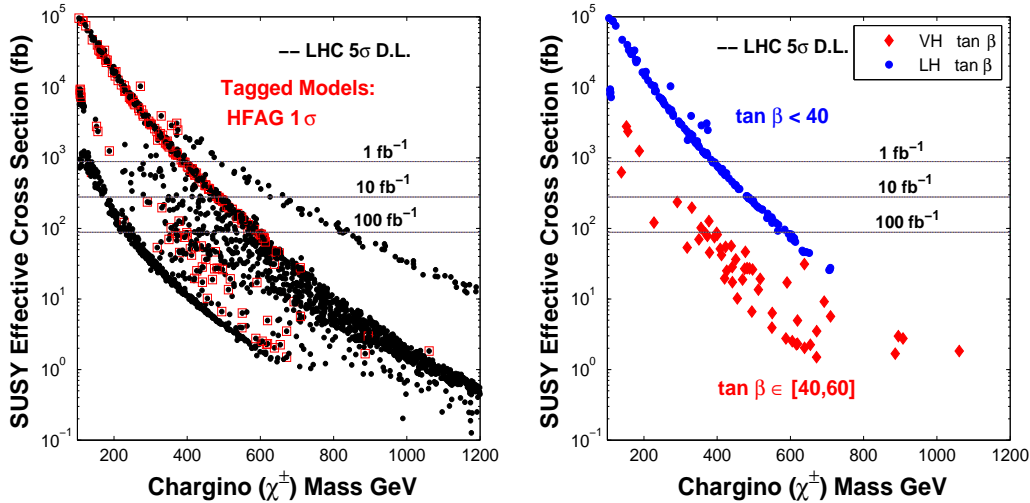}
\caption{ 
Total SUSY signatures at $\sqrt{s} = 14 ~ \rm TeV$ analyzed with the SUSY detector cuts.
Left: (circle,black) All models up to $1.2$ TeV in the chargino masses, and (red,boxed)
models within 1$\sigma$ corridor of the HFAG value.    
Right: Models separated out in $\tan \beta$ that lie within 1$\sigma$ corridor of the HFAG.
The dashed lines indicate the backgrounds, 5$\sqrt{\rm SM}$, 
for different luminosities. }
\label{h}
\end{figure}

\noindent

{\bf  Production and Signatures of  Sparticles}: 
In the following, we focus our analysis on the models that are favored
by the $b\to s\gamma$ constraint, namely, models that fall within  a 
1$\sigma$ 
corridor around the HFAG value.  
We discuss here the signatures of the $2 \to 2$ SUSY processes. 
In the analysis we use 
SuSpect to create a  SUSY Les Houches Accord (SLHA)  \cite{SKANDS} file 
which is then used as an input for 
 PYTHIA  \cite{PYTHIA} which computes the production cross 
sections and branching fractions, and for PGS  \cite{PGS} which simulates the 
LHC detector effects. The Level $1$  (L1) trigger cuts  
based on  the Compact Muon Solenoid detector specifications 
 \cite{CMS} are employed to analyze the LHC events. 
For our analysis of    sparticles, 
we further impose the post trigger detector cuts as follows:
We only select photons, electrons, and muons 
that have transverse momentum $P_T>10$ GeV and pseudorapidity $|\eta|<2.4$, 
taus jets that have $P_T>10$ GeV and $|\eta|<2.0$, and other hadronic jets that have 
$P_T>60$ GeV and $|\eta|<3$. 
We also require a large missing energy, $\not\!\!{P_T}>200$ GeV and
at least two jets in an event to further suppress the Standard Model (SM) background. 
We will refer this set of cuts as ``SUSY detector cuts'' in the following analysis 
(for other recent works on signature analysis of SUGRA models see  \cite{sugrasigs}).

We analyze the total number of events arising from the models 
in a 1$\sigma$ corridor around the HFAG results out of the 3$\sigma$ corridor  
using the SUSY detector cuts. The effective SUSY cross sections 
are then translated from the total number of events which are exhibited in Fig.(\ref{h}).  
One finds that the models with low values of $\tan\beta$  have strong SUSY signals
since these models tend to have a light sparticle spectrum, 
e.g., a  light stau,a light stop or even a light gluino. 
Most of the  LH $\tan\beta$ models discussed
 above can be probed at the LHC at $100$ fb$^{-1}$ of 
integrated luminosity.
It is found that the HFAG 1$\sigma$ constraint places a limit on
the chargino mass of about 800 GeV  for detectable models.   
We note that different models with different mass hierarchical 
spectra can have distinct SUSY signatures. For instance, 
models that have a light stau are rich in lepton signals, while 
models with a light stop tend to produce 
a high multiplicity of jet signals  . 
Thus the search strategies for  new physics at the LHC 
for such models are quite different, and a well designed search 
technique for every specific model will surely further improve 
the discovery reach. 
Nevertheless, the models 
that have low values of $\tan\beta$  have strong SUSY production 
cross sections, and  can be probed at the LHC. 
From the SUSY production analysis, one also finds that most of 
the VH $\tan\beta$ models have much smaller SUSY cross sections.

\begin{figure}[h]
\includegraphics[width=8cm,height=7cm]{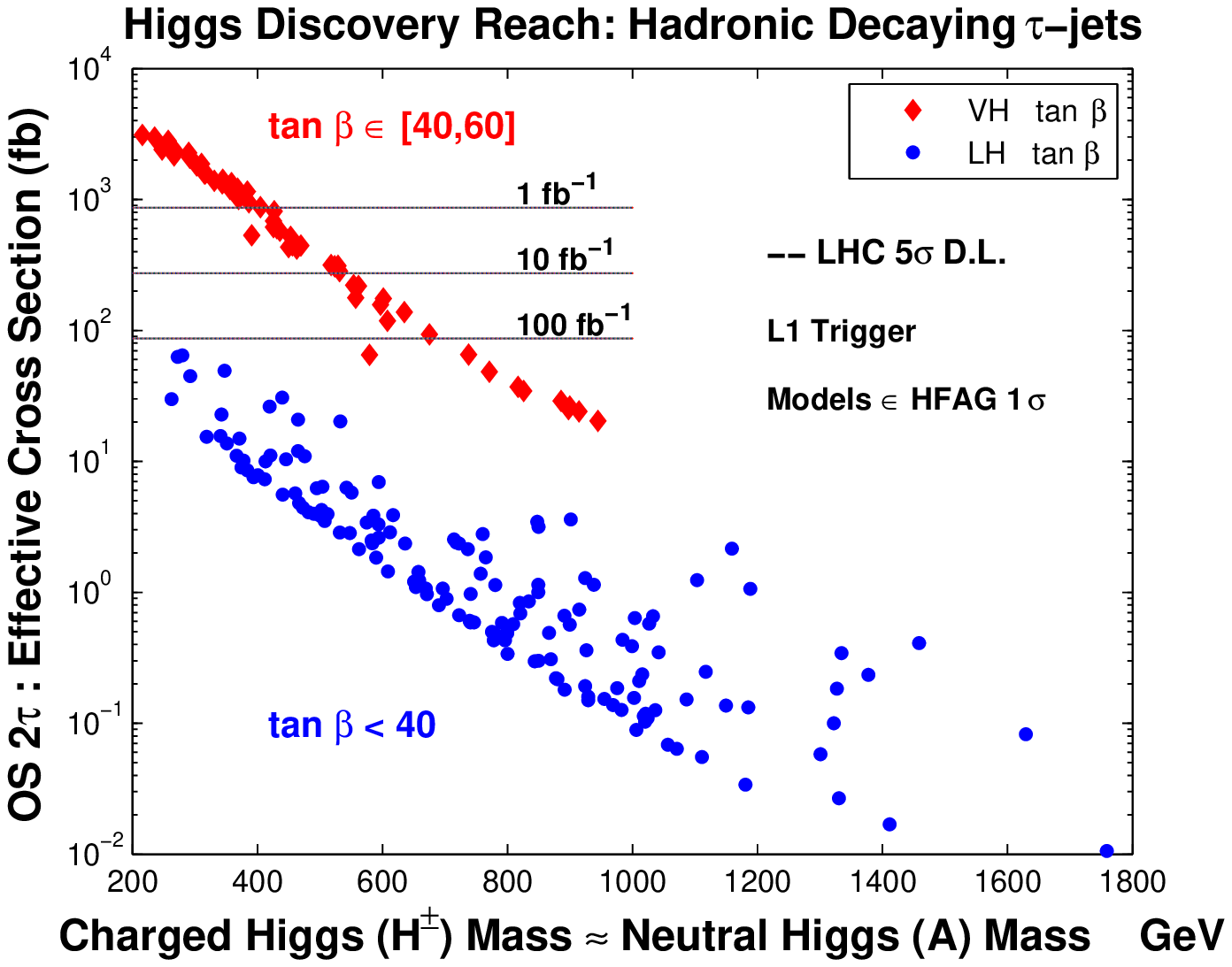}
\includegraphics[width=8cm,height=7cm]{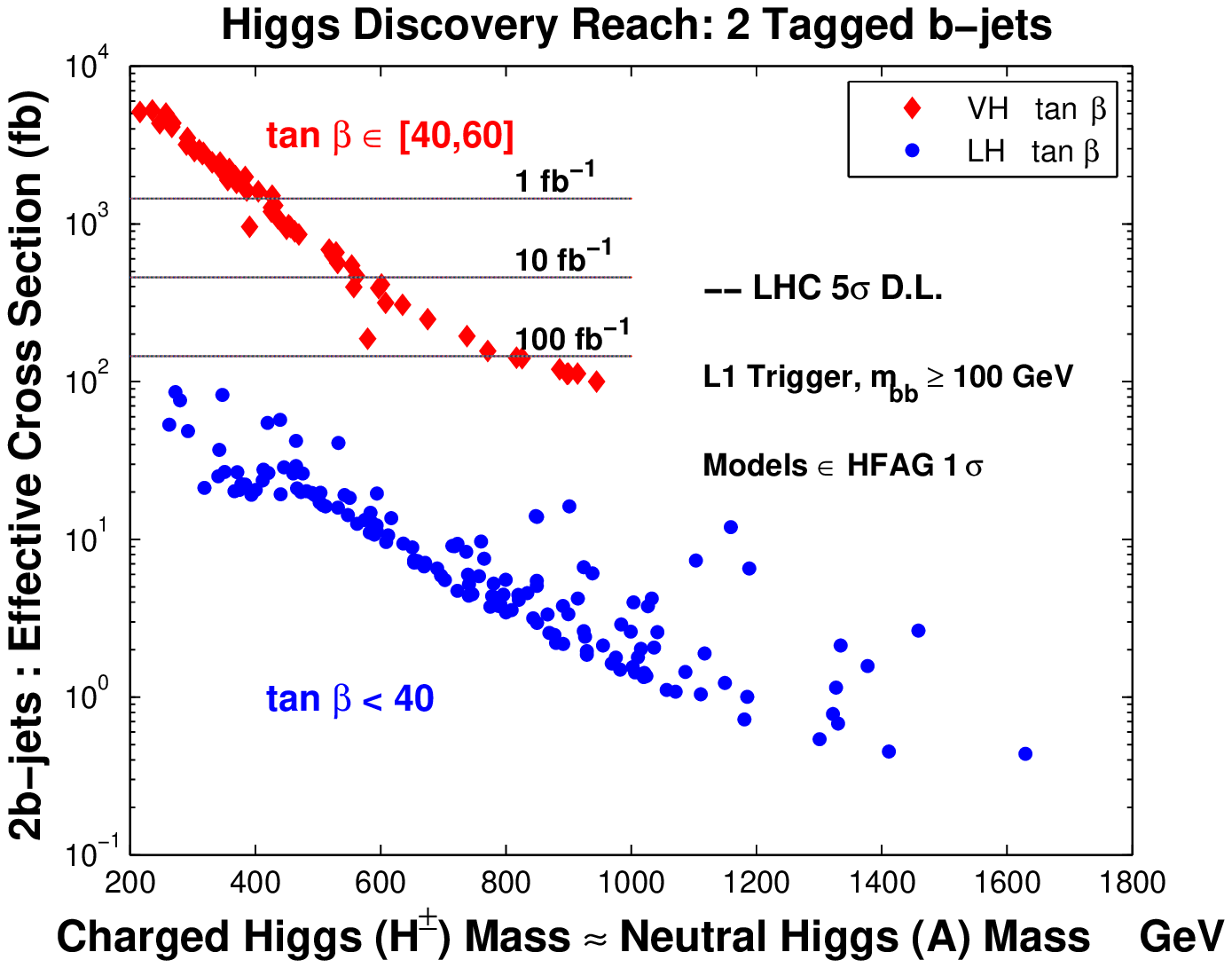}
\caption{ 
Higgs signatures with b-tagged jets and hadronic $\tau$-jets at $\sqrt{s} = 14 ~ \rm TeV$.
The dashed lines indicate the backgrounds, 5$\sqrt{\rm SM}$, 
for different luminosities. 
} 
\label{fig:sighiggs}
\end{figure}

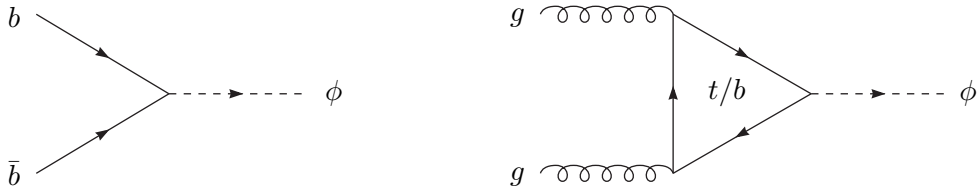
\begin{figure}
\begin{center}
\fcolorbox{white}{white}{
  \begin{picture}(400,80) (170,-140)
    \SetWidth{0.5}
    \SetColor{Black}
    \Line[arrow,arrowpos=0.5,arrowlength=3,arrowwidth=1,arrowinset=0.1](200,-70)(250,-100)
    \Line[arrow,arrowpos=0.5,arrowlength=3,arrowwidth=1,arrowinset=0.1](200,-130)(250,-100)
    \Line[dash,dashsize=3,arrow,arrowpos=0.5,arrowlength=3,arrowwidth=1,arrowinset=0.1](250,-100)(300,-100)
    \Text(190,-75)[lb]{$b$}
    \Text(190,-135)[lb]{$\bar b$}
    \Text(310,-105)[lb]{$\phi$}
    \Gluon(390,-70)(440,-70){3}{5}
    \Gluon(390,-130)(440,-130){3}{5}
    \Line[arrow,arrowpos=0.5,arrowlength=3,arrowwidth=1,arrowinset=0.1](440,-70)(492,-100)
    \Line[arrow,arrowpos=0.5,arrowlength=3,arrowwidth=1,arrowinset=0.1](492,-100)(440,-130)
    \Line[arrow,arrowpos=0.5,arrowlength=3,arrowwidth=1,arrowinset=0.1](440,-130)(440,-70)
    \Line[dash,dashsize=3,arrow,arrowpos=0.5,arrowlength=3,arrowwidth=1,arrowinset=0.1](492,-100)(542,-100)
    \Text(380,-75)[lb]{$g$}
    \Text(380,-135)[lb]{$g$}
    \Text(455,-105)[lb]{$t/b$}
    \Text(550,-105)[lb]{$\phi$}
    \end{picture}
}
\end{center}
\caption{Dominant leading order Higgs production diagrams via bottom quark annihilation and 
gluon fusion.  For large $\tan\beta$ the bottom quark annihilation can dominate 
the gluon fusion process in some regions of the parameter space.  }
\label{fig:higgsdiag}
\end{figure}

\noindent
{\bf LHC Signatures in Higgs Production}:
We discuss here the signatures  of the Higgs bosons in MSSM 
(the CP-even Higgs $H^0$, the CP-odd Higgs $A^0$, and the 
charged Higgs $H^\pm$)
for the models that are within the 1$\sigma$ corridor of the HFAG value.  
Specially, we are interested in the parameter region where 
the $\tan\beta$ value becomes very large. 
As discussed previously, the VH $\tan\beta$ models  within the 1 $\sigma$ corridor of HFAG 
have less promising SUSY signals. 
However, the Higgs production can be much enhanced at  very high  $\tan\beta$.
The dominant processes that lead to the production of the MSSM Higgs bosons at the LHC   
for $\tan\beta \gg 1$ are the bottom quark annihilation process and the gluon 
fusion process  \cite{Djouadi:1991tka} shown in Fig.(\ref{fig:higgsdiag}) 
along with associated production processes with bottom quarks.

\begin{figure}[t]
\includegraphics[width=7cm,height=6cm]{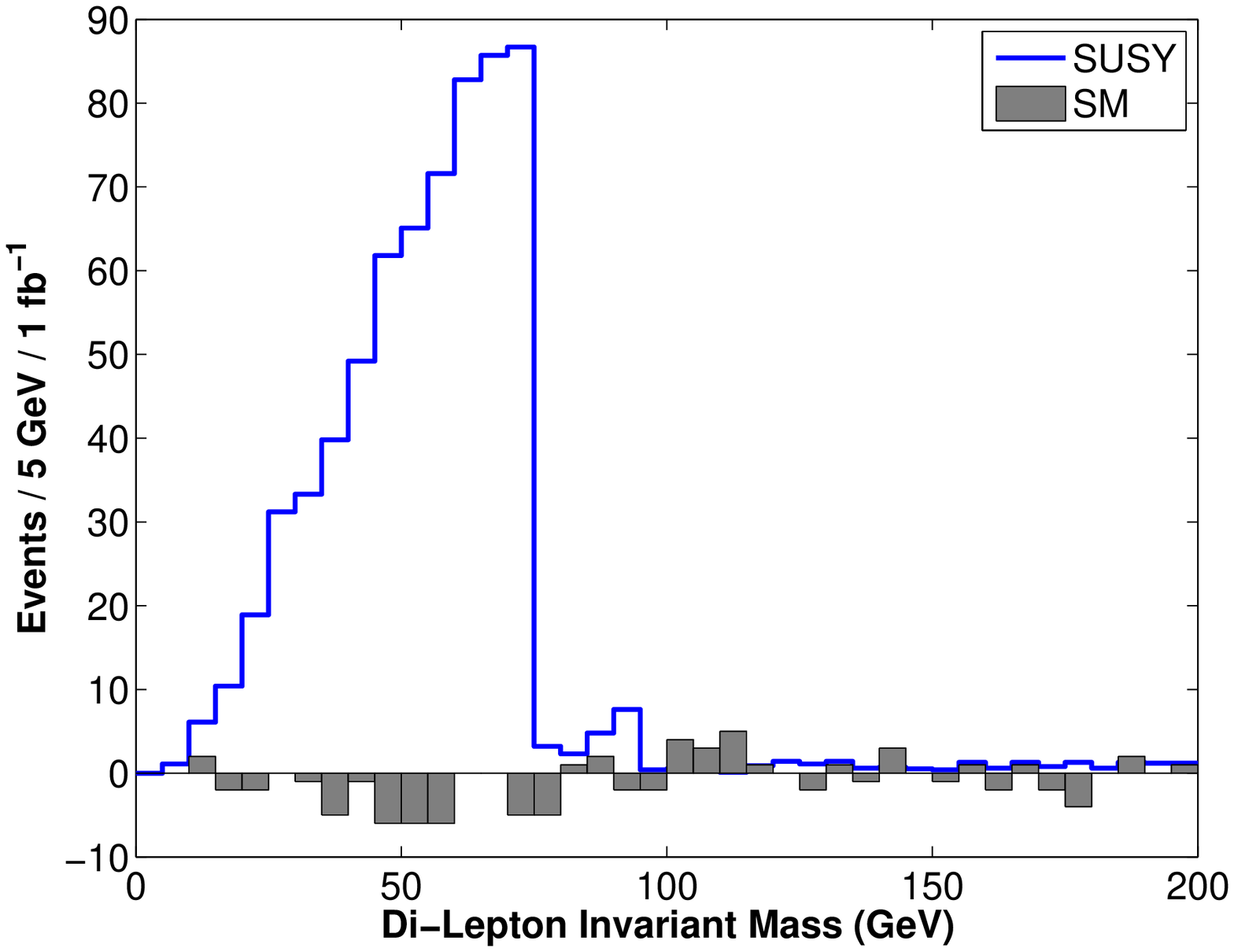}
\includegraphics[width=7cm,height=6cm]{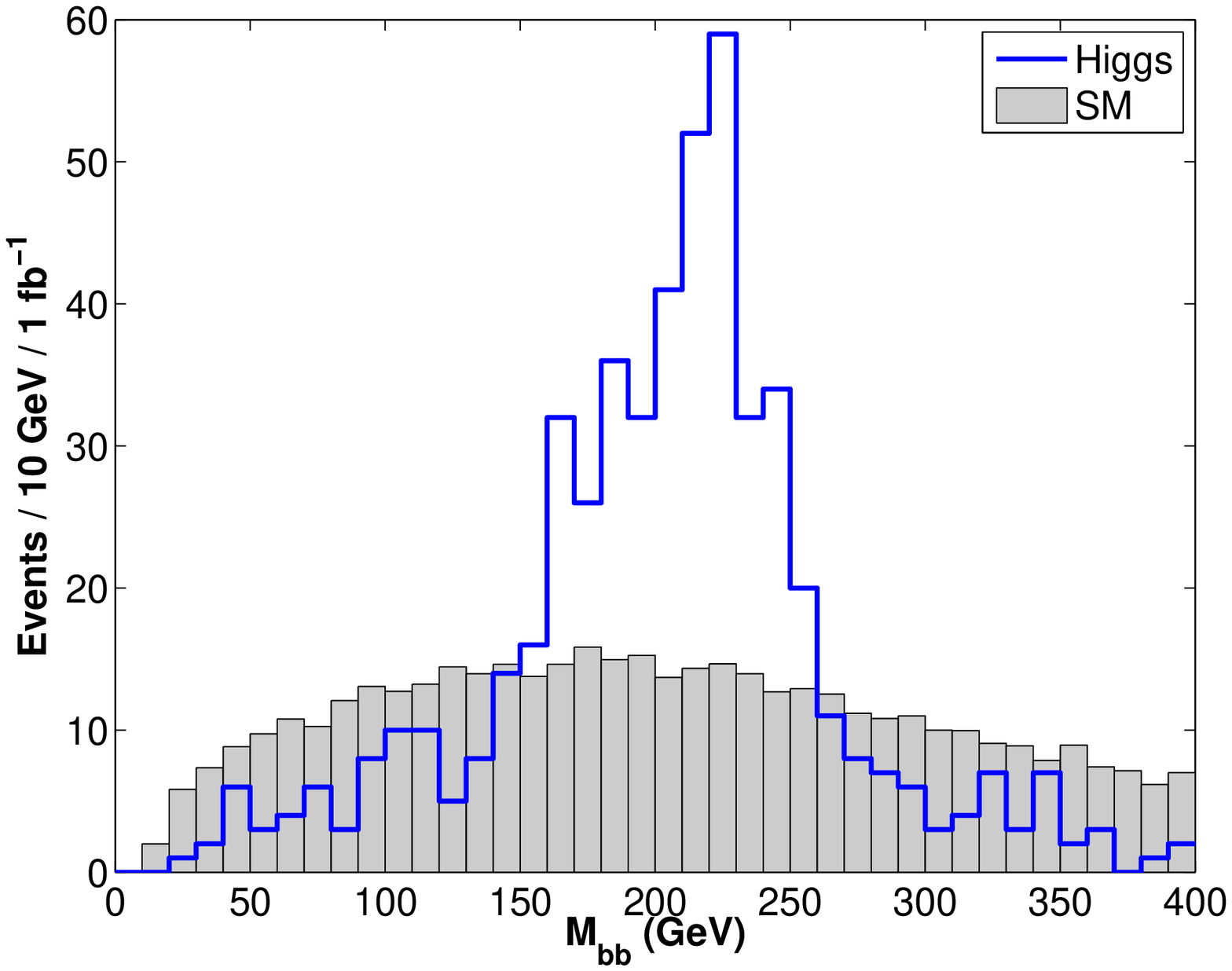}
\caption{
Invariant mass distributions for SUSY and Higgs productions for two different models: 
($m_0$, $m_{1/2}$, $A_0$, $\tan\beta$, sign($\mu$))=(70.4, 243.2, 685.6, 11, 1) (left panel); 
(1533.8, 216.4, 1750.3, 53.8, 1) (right panel) where all masses are in GeV.
Left: The opposite sign di-lepton with flavor subtraction  
($e^+e^-+\mu^+\mu^--e^+\mu^--\mu^+e^-$), for the model that fall 
within $1\sigma$ for both $b\to s\gamma$ and $g_\mu-2$  constraints. In this 
model, $M_{\na}=$ 93 GeV and $M_{\nb}=$  168 GeV. The ending edge of 
the distribution  indicates the mass difference ($M_{\nb}-M_{\na}$). 
Analysis is done with SUSY detector cuts. 
Shaded regions are the background $N_{\rm SM}$. 
Right: Reconstruction of the two hardest b-tagged jets in 3 b-jets events  
of Higgs productions for the model that satisfies the HFAG $1\sigma$. 
The peak indicates the position of the Higgs boson mass. 
L1 trigger cuts are employed. 
Shaded regions are the background $\sqrt{N_{\rm SM}}$. 
} 
\label{fig:mll}
\end{figure}

In our analysis, we focus on the hadronic $\tau$ and  jet production 
with bottom quark tagging, since the 
$b\bar b$ and $\tau^+\tau^-$   modes are the dominant 
decays of the MSSM Higgs bosons at large $\tan\beta$. 
We analyze the opposite sign (OS) di-tau signature and 
the 2b-jets signature using the L1 trigger cuts. 
For the 2b jet signatures, we also require the reconstructed invariant mass 
of these two b-tagged jets to be larger than 100 GeV.  
An analysis of the signatures for these 
models reveals the $2\tau$ jet and the $2b$  jet channels 
to be two of the optimal channels for the discovery of 
the Higgs bosons as shown in Fig.(\ref{fig:sighiggs}). 
It is found that the HFAG 1$\sigma$ constraint places a limit on
the charged Higgs mass about 1 TeV for the detectable models which 
can be probed with $L=$ 100 fb$^{-1}$ or so at LHC. We note that 
for the region $\tan \beta < 40$, one needs much more luminosity to 
observe discoverable events from the Higgs production, and some 
of the models in this region may be even beyond the LHC reach in the Higgs production. 
Thus the VH $\tan\beta$ models are  discoverable via 
Higgs production modes, while many of them have undetectable signals via 
sparticles productions. Thus the more optimal channels to discover supersymmetry in 
these VH $\tan\beta$ models  arise from Higgs production signals
as they produce larger event rates than the event rates from SUSY 
production processes with R-parity odd particles. 
The associated production in which the Higgs bosons 
are produced along with one or two bottom quarks in the 
final states can be very useful for suppressing further the 
SM background 
 \cite{Balazs:1998nt,Miller:1999bm,Campbell:2002zm,Maltoni:2003pn,Harlander:2003ai,Carena:2005ek,Belyaev:2005ct}. 
One example of the associated production with one additional bottom quark in the 
final state is given in Fig.(\ref{fig:mll}). 
For the hadronic $\tau$ jets signature, we utilize both the 1-prong and 
the 3-prong hadronic $\tau$-jets  \cite{Guchait:2006jp} in our analysis. 
We note that the leptonic decay modes of the 
$\tau$ lepton and a combined analysis of leptonic and hadronic 
decays may yield an even better discovery reach 
 \cite{Dawson:2004nv,Kao:2007qw, ATLAS}.

\begin{figure}[h]
\includegraphics[width=7cm,height=6cm]{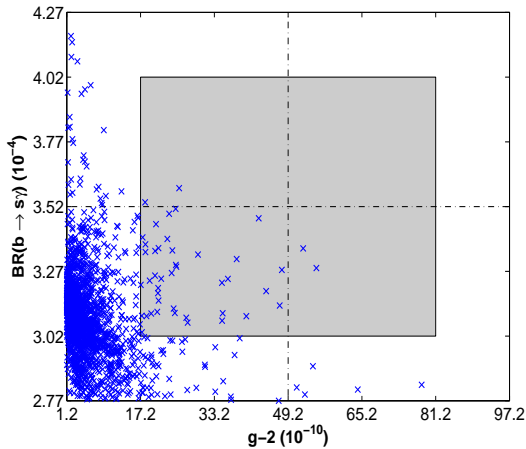}
\caption{
Combined analysis with $b\to s\gamma$ and $g_\mu-2$ constraints. Shaded 
regions are the $2\sigma$ corridors from both constraints.  
} 
\label{fig:g2b2sg}
\end{figure}

{\bf Complementarity of  Signatures from Sparticle Decays and from Higgs Decays}:
Before discussing the issue of complementarity  we discuss first the more stringent constraints 
arising from the recent revised analyses of 
$g_{\mu}-2$ which seem to converge   \cite{Davier:2009zi} 
towards a $3\sigma$  deviation from the standard model value.
 Fig.(\ref{fig:g2b2sg}) illustrates  a $2\sigma$ corridor around the central 
values of $\delta a_{\mu}$ and of the HFAG value of ${\mathcal Br}(b\to  s\gamma)$.
The analysis of Fig.(\ref{fig:g2b2sg})  shows that the  parameter space of allowed models
is drastically reduced.  A model point   from the allowed set of models is discussed in further
detail in the context of complementarity below. 

\begin{figure}[h]
\includegraphics[width=8cm,height=6cm]{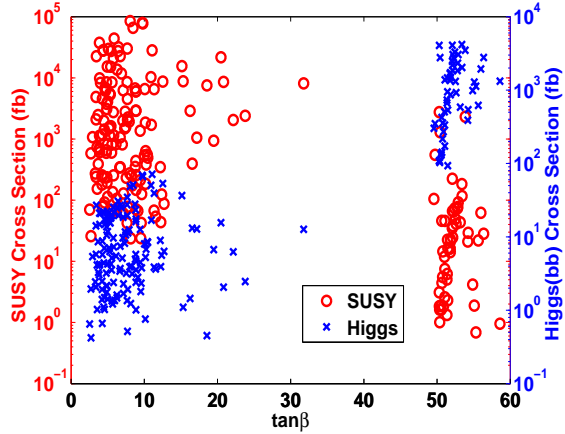}
\caption{
A plot of both SUSY signatures and Higgs signatures for the models 
that fall within the 1$\sigma$ corridor around the HFAG value. 
The figure shows complementarity and inversion, in the sense that at
low $\tan\beta$ sparticle production cross sections  dominate while 
at  high $\tan\beta$ the Higgs  production cross sections dominate.
} 
\label{fig:susyhiggs}
\end{figure}

Next we point out a  complementarity that exists between two main 
types of processes  in the 
production and decays of new particles expected at the LHC.  The first of the  main types
consists of those production processes which have in their final decay products an even number 
of massive LSPs (each an R-parity odd particle).  These arise from the production and 
subsequent decay of an even number of R parity odd particles (due to R parity 
conservation)  such as  pairs of squarks or gauginos or both at the LHC. 
These processes are characterized by a large missing energy since the final
states have at least two or more LSPs,  which for the models considered are 
the lightest neutralinos.  Thus here a larger missing 
transverse momentum is the smoking gun signature for the SUSY productions. 

The second type of processes are those which do not contain  pairs of 
LSPs and thus there is far less missing energy associated with these events. 
Such events are expected to arise from the production of the Higgs
bosons where the dominant decay products are largely $b\bar b$ and
$\tau^+\tau^-$. The  signals arising  from the Higgs decays typically 
suffer from a large QCD background, since the $\not\!\!{P_T}$ cut 
technique cannot be employed here which is efficient in suppressing 
the background for SUSY production.
However, for models with very high $\tan\beta$, 
the Higgs production 
is enhanced and  such model points can yield  signals which can be 
discriminated from the QCD background. Thus we see that there 
is a complementarity between the signatures arising from the
production and decay of the SUSY particles and from the Higgs particles, 
and this complementarity is exhibited in Fig.(\ref{fig:susyhiggs}). 
Indeed for models with small $\tan\beta$, missing energy continues 
to be a dominant signal while for models with  very high  $\tan\beta$ 
and within $1\sigma$ corridor of HFAG value of $b\to s\gamma$  
the Higgs  production and decay into $b\bar b$ and $\tau^+\tau^-$ 
can provide  signatures which can supersede the signatures  from
sparticle production  for the discovery of supersymmetry at the LHC.

A more detailed signal analysis on SUSY production and on Higgs production 
is given in Fig.(\ref{fig:mll}). In the left panel of Fig.(\ref{fig:mll}), 
the model considered is the one where stau is the NLSP and it shows a 
strong SUSY production signal which is rich in lepton final states. 
The lightest sparticles in this model besides the LSP are 
$\tilde\tau$, $\tilde\ell_R$, so in the cascade decays of heavier 
gauginos, these sleptons can appear in the intermediate steps, for instance,  
$\tilde\chi^0_2$ decays predominantly via BR$(\tilde\tau+\tau)\sim$ 70\% 
and BR$(\tilde\ell_R+\ell)\sim$ 20\%. The produced sleptons further decay 
into the LSP plus one lepton. Thus the reconstruction of the di-lepton events 
indicate the mass relations 
between the gauginos in the cascade decay chain due to the missing energy 
carried away by the LSP. 
In contrast, the invariant mass of the b-tagged jets from the Higgs production  
gives rise to a resonance which points to the actual value of the Higgs boson mass
as exhibited by the right panel of Fig.(\ref{fig:mll}).
As stated in the caption of  Fig.(\ref{fig:mll}) the cuts used in the left panel are the
SUSY detector cuts which are discussed in the first paragraph of the section on 
"Production and signatures of sparticles". The right panel of  Fig.(\ref{fig:mll}) is analyzed with the 
standard L1 trigger cuts in PGS. The 
background in the left panel of Fig.(\ref{fig:mll}) is suppressed by using 
 flavor subtraction in reconstructing the dilepton events. 
 For the right panel, the background is suppressed by 
 reconstructing the two hardest b-tagged jets in the 3b events which can arise in the associated production modes of Higgs bosons. The associated production where the Higgs bosons are produced along with additional b-tagged jets 
 is instrumental in suppressing the background. 
  \\
 
\noindent
{\bf Conclusion:} 
${\mathcal Br}(b\to s\gamma)$ in the standard model re-evaluated by the inclusion of NNLO corrections 
falls  below the central HFAG value by about $(1-1.5)\sigma$ hinting at a positive contribution
from the supersymmetric sector.  The obvious candidate for a positive contribution is  the charged
Higgs exchange. On the other hand the chargino exchange  contributions can produce either a
positive or a negative contribution.  Further, the recent re-evaluations of the $g_{\mu}-2$ indicate
about a $3.1\sigma$  deviations from the standard model pointing to the possibility of a light chargino. 
A detailed investigation of the parameter space of supergravity models reveals that most model points 
that satisfy both the ${\mathcal Br}(b\to s\gamma)$ and the $g_{\mu}-2$ constraints produce 
both a light charged Higgs and a light chargino with a cancellation between the charged Higgs 
loops and the chargino loops indicating the existence of some of the sparticle masses, 
specifically the chargino, charged Higgs and the stop being light. 
We have emphasized the importance of studying simultaneously sparticle and Higgs production.
The implications of these results for early SUSY discovery at the LHC were discussed and 
it is shown that some of the sparticles can be discovered in  runs with low luminosity. \\

\noindent
{\em Acknowledgments}:  
This research is  supported in part by NSF grant PHY-0653342 (Stony Brook),
DOE grant  DE-FG02-95ER40899 (MCTP), and NSF grant PHY-0757959 (Northeastern University). 
NC and ZL would like to thank Alex Mitov and Robert Shrock for helpful discussions. 



\begin{thebibliography}{999}

  \bibitem{Misiak:2006zs}
  M.~Misiak {\it et al.},
  Phys.\ Rev.\ Lett.\  {\bf 98} (2007) 022002.

\bibitem{Barberio:2008fa}
  E.~Barberio {\it et al.}  [Heavy Flavor Averaging Group],
  arXiv:0808.1297 [hep-ex].

\bibitem{bbmr}
  S.~Bertolini, F.~Borzumati, A.~Masiero and G.~Ridolfi,
  Nucl.\ Phys.\  B {\bf 353}, 591 (1991).

\bibitem{Nath:1994tn}
  P.~Nath and R.~L.~Arnowitt,
  Phys.\ Lett.\  B {\bf 336}, 395 (1994);
 F.~Borzumati, M.~Drees and M.~M.~Nojiri,
  Phys.\ Rev.\  D {\bf 51}, 341 (1995).

\bibitem{susybsgamma}
 G.~Degrassi, P.~Gambino and G.~F.~Giudice,
  JHEP {\bf 0012} (2000) 009;
  F.~Borzumati, C.~Greub, T.~Hurth and D.~Wyler,
  Phys.\ Rev.\  D {\bf 62}, 075005 (2000)
D.~A.~Demir and K.~A.~Olive,
  Phys.\ Rev.\  D {\bf 65}, 034007 (2002);
   A.~J.~Buras et.al.,
  Nucl.\ Phys.\  B {\bf 659} (2003) 3;
   M.~E.~Gomez, T.~Ibrahim, P.~Nath and S.~Skadhauge,
  Phys.\ Rev.\  D {\bf 74} (2006) 015015;
   G.~Degrassi, P.~Gambino and P.~Slavich,
  Phys.\ Lett.\  B {\bf 635} (2006) 335.

\bibitem{Hewett:1992is}
  J.~L.~Hewett,
  Phys.\ Rev.\ Lett.\  {\bf 70}, 1045 (1993);
  V.~D.~Barger, M.~S.~Berger and R.~J.~N.~Phillips,
  Phys.\ Rev.\ Lett.\  {\bf 70}, 1368 (1993).

\bibitem{Garisto:1993jc}
  R.~Garisto and J.~N.~Ng,
  Phys.\ Lett.\  B {\bf 315}, 372 (1993).


\bibitem{Davier:2009zi}
  M.~Davier, A.~Hoecker, B.~Malaescu, C.~Z.~Yuan and Z.~Zhang,
  arXiv:0908.4300 [hep-ph].

\bibitem{yuan}
  T.~C.~Yuan, R.~L.~Arnowitt, A.~H.~Chamseddine and P.~Nath,
  Z.\ Phys.\  C {\bf 26}, 407 (1984);
 D.~A.~Kosower, L.~M.~Krauss and N.~Sakai,
  Phys.\ Lett.\  B {\bf 133}, 305 (1983);
 J.~L.~Lopez, D.~V.~Nanopoulos and X.~Wang,
  Phys.\ Rev.\  D {\bf 49}, 366 (1994);
 J.~L.~Lopez, D.~V.~Nanopoulos and X.~Wang,
  Phys.\ Rev.\  D {\bf 49}, 366 (1994).
 T.~Moroi,
  Phys.\ Rev.\  D {\bf 53}, 6565 (1996).

\bibitem{Czarnecki:2001pv}
A.~Czarnecki and W.~J.~Marciano,
Phys.\ Rev.\ D {\bf 64}, 013014 (2001).

\bibitem{Chattopadhyay:2001vx}
  U.~Chattopadhyay and P.~Nath,
  Phys.\ Rev.\ Lett.\  {\bf 86}, 5854 (2001).

\bibitem{Everett:2001tq}
  L.~L.~Everett, G.~L.~Kane, S.~Rigolin and L.~T.~Wang,
  Phys.\ Rev.\ Lett.\  {\bf 86}, 3484 (2001).
  
\bibitem{fm}
J.~L.~Feng and K.~T.~Matchev, Phys.\ Rev.\ Lett.\  {\bf 86}, 3480 (2001);
 E.~A.~Baltz and P.~Gondolo, Phys.\ Rev.\ Lett.\  {\bf 86}, 5004 (2001);
T. Ibrahim, U. Chattopadhyay and P. Nath, Phys. Rev. {\bf D64}, 
016010(2001); J. Ellis, D.V. Nanopoulos, K. A. Olive, Phys.\ Lett.\ B {\bf 508}, 65 (2001);
R. Arnowitt, B. Dutta, B. Hu, Y. Santoso, 
Phys.\ Lett.\ B {\bf 505}, 177 (2001);
S. P. Martin, J. D. Wells, Phys.\ Rev.\ D {\bf 64}, 035003 (2001);
H. Baer, C. Balazs, J. Ferrandis, X. Tata, 
Phys.Rev.{\bf D64}: 035004, (2001). 

\bibitem{msugra}
 A.~H.~Chamseddine, R.~Arnowitt and P.~Nath,
  Phys.\ Rev.\ Lett.\  {\bf 49}, 970 (1982);
 P.~Nath, R.~Arnowitt and A.~H.~Chamseddine,
  Nucl.\ Phys.\  B {\bf 227}, 121 (1983);
L.~J.~Hall, J.~D.~Lykken and S.~Weinberg,
  Phys.\ Rev.\  D {\bf 27} (1983) 2359.

\bibitem{Feldman:2007zn}
  D.~Feldman, Z.~Liu and P.~Nath,
  Phys.\ Rev.\ Lett.\  {\bf 99}, 251802 (2007);
  D.~Feldman, Z.~Liu and P.~Nath,
  JHEP {\bf 0804}, 054 (2008).



\bibitem{Komatsu:2008hk}
  E.~Komatsu {\it et al.}  [WMAP Collaboration],
  Astrophys.\ J.\ Suppl.\  {\bf 180}, 330 (2009).

\bibitem{Spergel:2006hy}
  D.~N.~Spergel {\it et al.},
  astro-ph/0603449.

\bibitem{2007kv}
  T.~Aaltonen {\it et al.}  [CDF Collaboration],
  Phys.\ Rev.\ Lett.\  {\bf 100}, 101802 (2008).

\bibitem{Djouadi:2006be}
  A.~Djouadi, M.~Drees and J.~L.~Kneur,
  JHEP {\bf 0603}, 033 (2006).

\bibitem{Djouadi:1991tka}
  A.~Djouadi, M.~Spira and P.~M.~Zerwas,
  Phys.\ Lett.\  B {\bf 264}, 440 (1991);
 M.~Spira, A.~Djouadi, D.~Graudenz and P.~M.~Zerwas,
  Nucl.\ Phys.\  B {\bf 453} (1995) 17.




\bibitem{Dawson:2005vi}
  S.~Dawson, C.~B.~Jackson, L.~Reina and D.~Wackeroth,
  Mod.\ Phys.\ Lett.\  A {\bf 21}, 89 (2006).
  
 


  
\bibitem{lephiggs}
The LEP Higgs Working Group, LHWG-Note 2004-01 (August 2004). 
   


\bibitem{MICRO}
  G.~Belanger, F.~Boudjema, A.~Pukhov and A.~Semenov,
  Comput.\ Phys.\ Commun.\  {\bf 176}, 367 (2007).


\bibitem{SUSPECT}
  A.~Djouadi, J.~L.~Kneur and G.~Moultaka,
  Comput.\ Phys.\ Commun.\  {\bf 176}, 426 (2007).

\bibitem{Degrassi:2007kj}
  G.~Degrassi, P.~Gambino and P.~Slavich,
  Comput.\ Phys.\ Commun.\  {\bf 179}, 759 (2008).


\bibitem{Feldman:2007fq}
  D.~Feldman, Z.~Liu and P.~Nath,
  Phys.\ Lett.\  B {\bf 662}, 190 (2008).


\bibitem{Barenboim:2007sk}
  G.~Barenboim, P.~Paradisi, O.~Vives, E.~Lunghi and W.~Porod,
  JHEP {\bf 0804}, 079 (2008).
  
\bibitem{Dudley:2009zi}
  B.~Dudley and C.~Kolda,
  arXiv:0901.3337 [hep-ph].

\bibitem{SKANDS}
  P.~Skands {\it et al.},
  JHEP {\bf 0407}, 036 (2004).

\bibitem{PYTHIA}
  T.~Sjostrand, S.~Mrenna, P.~Skands,
  JHEP {\bf 0605}, 026 (2006).


\bibitem{PGS}
\url{http://www.physics.ucdavis.edu/~conway/research/software/pgs/pgs4-general.htm}

\bibitem{CMS}
CMS Collaboration,
CERN/LHCC 2006-001 (2006).

\bibitem{sugrasigs}
  R.~Arnowitt {\it et al.},
  Phys.\ Lett.\  B {\bf 649}, 73 (2007);
  U.~Chattopadhyay, D.~Das, A.~Datta and S.~Poddar,
  Phys.\ Rev.\  D {\bf 76}, 055008 (2007);
  H.~Baer, A.~Mustafayev, E.~K.~Park and X.~Tata,
  JHEP {\bf 0805}, 058 (2008);
    D.~Feldman, Z.~Liu and P.~Nath,
  Phys.\ Rev.\  D {\bf 78}, 083523 (2008);
  %
  Phys.\ Rev.\  D {\bf 80}, 015007 (2009);
B.~Altunkaynak, P.~Grajek, M.~Holmes, G.~Kane and B.~D.~Nelson,
 JHEP {\bf 0904}, 114 (2009);
  D.~Feldman, Z.~Liu, P.~Nath and B.~D.~Nelson,
  Phys.\ Rev.\  D {\bf 80}, 075001 (2009);
  S.~Bhattacharya, U.~Chattopadhyay, D.~Choudhury, D.~Das and B.~Mukhopadhyaya,
  arXiv:0907.3428 [hep-ph];
    H.~Baer, V.~Barger, A.~Lessa and X.~Tata,
  arXiv:0907.1922 [hep-ph];
    D.~Feldman,
  arXiv:0908.3727 [hep-ph].


 


  

\bibitem{Balazs:1998nt}
  C.~Balazs, J.~L.~Diaz-Cruz, H.~J.~He, T.~M.~P.~Tait and C.~P.~Yuan,
  Phys.\ Rev.\  D {\bf 59}, 055016 (1999).
  

\bibitem{Miller:1999bm}
  D.~J.~Miller, S.~Moretti, D.~P.~Roy and W.~J.~Stirling,
  Phys.\ Rev.\  D {\bf 61}, 055011 (2000).
  
  

\bibitem{Campbell:2002zm}
  J.~M.~Campbell, R.~K.~Ellis, F.~Maltoni and S.~Willenbrock,
  Phys.\ Rev.\  D {\bf 67}, 095002 (2003).
 


\bibitem{Maltoni:2003pn}
  F.~Maltoni, Z.~Sullivan and S.~Willenbrock,
  Phys.\ Rev.\  D {\bf 67}, 093005 (2003).


\bibitem{Harlander:2003ai}
  R.~V.~Harlander and W.~B.~Kilgore,
  Phys.\ Rev.\  D {\bf 68}, 013001 (2003).
  
  
\bibitem{Belyaev:2005ct}
  A.~Belyaev, A.~Blum, R.~S.~Chivukula and E.~H.~Simmons,
  Phys.\ Rev.\  D {\bf 72}, 055022 (2005);
  U.~Aglietti {\it et al.},
  arXiv:hep-ph/0612172.

\bibitem{Carena:2005ek}
  M.~S.~Carena, S.~Heinemeyer, C.~E.~M.~Wagner and G.~Weiglein,
  Eur.\ Phys.\ J.\  C {\bf 45}, 797 (2006).
  
  


\bibitem{Guchait:2006jp}
  M.~Guchait, R.~Kinnunen and D.~P.~Roy,
  Eur.\ Phys.\ J.\  C {\bf 52}, 665 (2007).
  
 
\bibitem{Dawson:2004nv}
  S.~Dawson, D.~Dicus, C.~Kao and R.~Malhotra,
  Phys.\ Rev.\ Lett.\  {\bf 92}, 241801 (2004).
  

\bibitem{Kao:2007qw}
  C.~Kao, D.~A.~Dicus, R.~Malhotra and Y.~Wang,
  Phys.\ Rev.\  D {\bf 77}, 095002 (2008).
  
\bibitem{ATLAS}
C.~Lampen, ATL-PHYS-PROC-2009-122 (ATLAS Notes).


  

  

\end{thebibliography}
\end{document}